\documentclass[floatfix,aps,showpacs,twocolumn,nofootinbib,preprintnumbers]
{revtex4}

\begin{document}

\title{Lorentz invariance violation in top-down scenarios of ultrahigh
energy cosmic ray creation} 

\preprint{FERMILAB-Pub-03/183-A}

\author{James R. Chisholm}
\author{Edward W. Kolb}
\affiliation{Fermilab Astrophysics, Fermi National Accelerator Laboratory, 
Batavia, Illinois 60510-0500, and \\ Enrico Fermi Institute, University of Chicago, Illinois
60637}

\date{\today}

\begin{abstract} 
The violation of Lorentz invariance (LI) has been invoked in a number of ways 
to 
explain issues dealing with ultrahigh energy cosmic ray (UHECR) production and
propagation.  These treatments, however, have mostly been limited to examples
in the proton-neutron system and photon-electron system.
In this paper we show how a broader violation of Lorentz invariance would 
allow for a series of previously forbidden decays to occur, and how that 
could lead 
to UHECR primaries being heavy baryonic states or Higgs bosons.
\end{abstract}


\maketitle

\section{Introduction}

UHECRs are an enduring mystery.  Without the introduction of new
particles or interactions, evading the Greisen--Zatsepin--Kuz'min
(GZK) cutoff \cite{greisen, zatsepin} requires unidentified nearby
sources.  Even without the GZK cutoff, the ``bottom-up'' approach
faces the challenge of finding in nature an accelerator capable of
energies in excess of $10^{20}$eV.

``Top-down'' scenarios assume that the UHECRs result from the
fragmentation of a ultra-high energy hadronic jet produced by cosmic
strings \cite{HillSchramm} or by the decay of a supermassive particle
\cite{supermassive}.  In the supermassive particle (wimpzilla)
scenario, the UHECRs are of galactic origin, resulting from the decay
of relic supermassive ($M\gtrsim 10^{13}$GeV) particles.  Wimpzillas
can be produced copiously in the early universe \cite{wimpzillas},
thus solving the energy problem.  Since they would cluster in the dark
matter halo of our galaxy \cite{birkel,blasi, dick}, they also solve the
distance problem.  Detailed analysis of these decays, however, show
that at high energy top-down scenarios produce more photons than
protons \cite{sarkar,berezinsky,barbot} in the UHECR spectrum seen at Earth.

The top-down prediction of photon preponderance in UHECRs is the one
major problem in an otherwise simple explanation. Results from the 
Fly's Eye \cite{eye}, Haverah Park \cite{ave}, and
AGASA \cite{shinozaki} cosmic ray experiments all indicate that at
energies above around $10^{19}$ to $10^{20}$eV, protons are more
abundant than photons in UHECRs.  While photons are disfavored, it
is not possible to be sure that the primary is indeed a proton.

The idea of violating LI (see Ref.\ \cite{kostelecky}
for a broad overview) has recently been studied in the context of
UHECRs for proton decay \cite{CG2,dubovsky}, atmospheric shower
development \cite{vankov}, and modifications to the cutoffs of proton
and photon spectra due to cosmic background fields
\cite{sato,kifune,aloisio,bertolami,carvalho,AP2}.  As shown in Ref.\ \cite{CG1}, 
LI violation can also lead to vacuum photon decay, and in
general the decay of particles to more massive species.
We exploit this fact to show how particles produced in top-down decays can 
undergo an ``inverse cascade'' to produce superheavy UHECR primaries.

\section{Breaking the Law}

\subsection{Modified Dispersion Relations}

There are two approaches to breaking LI commonly used
in the literature.  The first \cite{JLM1,CG1,kifune,aloisio,dubovsky}
is generically to modify the dispersion relations for a particle of
mass $m$ and 3-momentum $p = |\bf{p}|$.  

The second method, which we will not utilize here, is instead to write
down a particle Lagrangian \cite{colladay} which includes Lorentz and
CPT violating terms.  A mapping of the Lagrangian terms to a
dispersion relation is non-trivial (see \cite{carvalho} for a simple
case); however, to first order the changes to the photon and electron
propagators induce shifts to $c_\gamma$ and $c_e$ \cite{CG2}.  Similarly, it 
was
shown by \cite{AP1} that loop quantum gravity effects produce modified
dispersion relations similar to those considered elsewhere and here.

The simplest way to break LI, as shown in \cite{CG2}, is
to write down a dispersion relation for a particle species $i$ as
\begin{equation}
\label{MAV}
E_i^2 = p_i^2 c_i^2 + m_i^2 c_i^4.
\end{equation}
This is changing the ``speed of light'' or Maximum
Attainable Velocity (MAV) for each particle to something slightly different 
than $c$.\footnote{Although factors of $c$ are explicitly included in equations,
we still take it to be dimensionless with $c_\gamma = c = 1$.}  
The MAVs for different particle species are assumed {\it a priori}
to be different in that and similar treatments \cite{JLM1}.  

In such a case, it is possible for previously unallowable reactions to occur,
such as $p \rightarrow n e^+ \nu_e$ \cite{CG2, dubovsky}, $\gamma \rightarrow
e^+ e^-$ \cite{CG1, JLM1} and $\gamma \rightarrow \nu \bar \nu$ \cite{CG1}.
A conclusion reached, using the first reaction, is that neutrons may make up
the dominant baryonic component of the UHECR primaries.  This, strictly, would
be true if one was only tuning the proton and neutron MAVs and leaving all 
others equivalent ($=c$).  On the other hand, keeping every MAV as 
individually tunable gives an overwhelming number of free parameters.

In
order to examine the consequences of varying all particle MAVs
using the fewest number of free parameters, we extend the
method used in Ref.\ \cite{aloisio} and write the dispersion relation as
\begin{eqnarray}
\label{dispersionrelation}
E^2 & = & p^2 c^2 + m^2 c^4 + p^2 c^2
f\left(p/Mc\right)\nonumber \\ & & + m^2 c^4
g\left(p/Mc\right) + m c^2 p c h\left(p/Mc\right) .
\end{eqnarray}
Here, $f(x), g(x)$, and $h(x)$ are dimensionless universal functions
having the property $f(0) = g(0) = h(0) = 0$, so that as $p/M
\rightarrow 0$ the normal dispersion relations are recovered.  
Here $M$ is the mass scale that determines the relative degree of Lorentz
violation; for our purposes we set it at the Planck mass  $M \approx 10^{19}$ 
GeV.
Expanding these functions in Taylor series about $x=0$ and keeping
terms of $O(p^2)$ and smaller gives
\begin{eqnarray}
\label{disprel2}
E^2 &=& m^2 c^4 + m c^2 p c \left(\frac{1}{2} g'(0)
\frac{m}{M}\right)\nonumber \\& & + p^2 c^2 \left(1 + \frac{1}{2} h'(0)
\frac{m}{M} + \frac{1}{6} g''(0) \frac{m^2}{M^2}\right) .
\end{eqnarray}
The $g'(0)$ term can be neglected for two reasons: at $p \ll mc$ it would be
experimentally detected as deviations in the non-relativistic kinetic energy;
for $p \gg mc$ it is 
negligible to the quadratic $p$ term.  Note that to this (lowest)
order, $f(x)$ has no effect.\footnote{Issues of photon dispersion
in the vacuum, as discussed in Refs.\ \cite{kifune,JLM2}, will not be
addressed here.}  For this section, we examine the cases of massless (photon) 
and
massive (electron) particles:
\begin{eqnarray}
\label{disp_g}
E_e^2 & = & m^2 c^4 + p_e^2 c^2 \left(1 + \frac{1}{2} h'(0) \frac{m}{M} +
\frac{1}{6} g''(0) \frac{m^2}{M^2}\right) \\
E_\gamma^2 & = & p_\gamma^2 c^2 .
\end{eqnarray}
We can now define
\begin{equation}
\label{defce}
c_e^2 = c^2 \left(1 + \frac{1}{2} h'(0) \frac{m}{M} + \frac{1}{6}
g''(0) \frac{m^2}{M^2}\right)
\end{equation}
\begin{equation}m_e^2 = m^2 \left(1 + \frac{1}{2} h'(0) \frac{m}{M} + 
\frac{1}{6} g''(0) \frac{m^2}{M^2}\right)^{-2}
\end{equation}
to then write the dispersion relation in the familiar form of Eqn.\ \ref{MAV}
as
\begin{equation}
\label{disp_e}
E_e^2 = m_e^2 c_e^4 + p_e^2 c_e^2 .
\end{equation}

\subsection{Example: Photon Decay}

In this section we consider the tree-level photon decay process $\gamma
\rightarrow e^+ e^-$, which is kinematically forbidden for $c_e = c$.
Following Ref.\ \cite{CG2}, we define the parameter $\delta_{\gamma e}$
as\footnote{This is identical to $\xi - \eta$ for $n=2$ in the
notation of Ref.\ \cite{JLM2}.}
\begin{equation}
\label{defdelta}
\delta_{\gamma e} = c^2 - c_e^2 .
\end{equation}
Using astrophysical constraints \cite{stecker,JLM2}, $\delta_{\gamma
e}$ can be limited to the range $ 10^{-16} > \delta_{\gamma e} > 
-10^{-17}$.
For $\delta_{\gamma e} > 0$, photon decay occurs above a photon energy
threshold given by
\begin{equation}
E_{th} = \frac{2 m_e c_e^2}{\sqrt{\delta_{\gamma e}}} c .
\end{equation}
This decay rate of was computed in Coleman \& Glashow (1997) as
\begin{equation}
\Gamma = \frac{1}{2} \alpha \left(\frac{\delta_{\gamma e}}{c_e^2}\right) 
E \left[1 - \left(E_{th}/E\right)^2 \right]^{3/2} .
\end{equation}

For decay product energy above about $2\times10^{20}$eV, photons will
outnumber protons, so we set $E_{th}$ to this value and
suggestively rewrite the decay rate above threshold as:
\begin{equation}
\frac{\Gamma}{\hbar c} \approx (1 \textrm{ km})^{-1} 
\left(\frac{E}{2\times10^{20} \textrm{ eV}}\right)
\left(\frac{\delta_{\gamma e}}{3\times10^{-29}}\right) .
\end{equation}

This would rule out all but a terrestrial source of UHE photons.  The
value $\delta_{\gamma e}=3\times10^{-29}$ is used is a lower limit; if
$\delta_{\gamma e}$ would be any lower the energy threshold would
become too high.  This means that only $\delta_{\gamma e} \in
(3\times10^{-29}, 10^{-16})$ will correct for the photon overabundance
in top-down scenarios.

Now we see what is required to have $\delta_{\gamma e}$ in the
necessary range to allow for this photon decay.  Using Eqs.\
(\ref{defce}) and (\ref{defdelta}), we can write
\begin{equation}
\delta_{\gamma e} = - \frac{1}{2} c^2 h'(0) \frac{m}{M} - 
\frac{1}{6} c^2 g''(0) \frac{m^2}{M^2} .
\end{equation}

Since the functions $g(x)$ and $h(x)$ are {\it a priori} arbitrary, so
are the values and signs of their derivatives.  It is argued in Ref.\
\cite{aloisio} that too strongly varying functions would be
unphysical, and that the derivatives should be of order
unity.\footnote{Note that if this holds also for $f(x)$, we erred in
not including the $f'(0)$ term in Eq.\ (\ref{disprel2}), as that
dominates over the $h'(0)$ term when $p \gtrsim mc$.  We will not
apologize much for this, as we are looking at the ``lowest order''
change in phenomenology (the change in the ``speeds of light'') and
not ``next order'' changes (such as vacuum photon dispersion).}
Adopting this, we can ignore the $g''(0)$ term in the above equation.
In this case, in order to have photon decay we need to require
$h'(0)$ negative.  Then $\delta_{\gamma e} \sim m c^2/2 M
\sim 3\times10^{-23}$ and $E_{th} = 2 m_e c_e^2 c
\sqrt{2 M/m c^2} = 2 \sqrt{2 m_e M} c^2 \approx
2\times10^{17}$eV.  This value of $\delta_{\gamma e}$ is
(logarithmically) in the middle of our allowable range.

Returning to our original impetus of addressing the feasibility of ``Top Down''
scenarios, we see that requiring $h'(0) \sim -1$ is sufficient to eliminate
any photons among the UHECR primaries.  This model, however, eliminates any
protons as well, which we will now address.

\subsection{Making the impossible possible}

Unlike previous investigations \cite{JLM2,CG2} where the MAV for each 
particle could be chosen independently, this is not
so in our case.  Since Eq.\ \ref{dispersionrelation} holds for all
types of particles, it is straightforward to see other effects of this
level of Lorentz invariance violation.  Using Eq.\
\ref{defce}, we can in general write for two particles $A$ and $X$
\begin{eqnarray}
\label{delta_ab}
\delta_{A X} & = & c_A^2 - c_X^2 = \frac{1}{2} h'(0) c^2 \frac{m_A - m_X}{M} 
\nonumber \\
             & \approx & \frac{(m_X - m_A)c^2}{2M},
\end{eqnarray}
where we have used $h'(0) \sim -1$. 

The consequences of this for particle decays are as follows, as first noted
by \cite{CG2}, and we reproduce their argument here. 

Consider the decay $A \rightarrow \{X\}$, where $\{X\}$ is a collection of 
massive particles.
We want to find the minimum possible energy $E_{min}$ 
where this
can occur.  We can always lower final state energy by removing transverse
components of momenta, so we're working in one dimension only.  We want to
minimize
\begin{equation}
E\left(\{p_X\}\right) = \sum_X \sqrt{m_X^2 c_X^4 + p_X^2 c_X^2}
\end{equation}
subject to the constraint
\begin{equation}
G\left(\{p_X\}\right) = \left(\sum_X p_X \right) - p_A = 0.
\end{equation}
The $\{p_X\}$ are variables, and $p_A$, $E_A > E_{min}$ are constants.
Using the method of Langrange multipliers, $E$ is minimized when
\begin{equation}
\frac{\partial E}{\partial p_X} = \lambda \frac{\partial G}{\partial p_X},
\end{equation}
which becomes
\begin{equation}
\lambda = \frac{p_X c_X^2}{E_X} = \frac{p_X c_X^2}{\sqrt{m_X^2 c_X^4 + p_X^2 c_X^2}}.
\end{equation}

Thus all products move with the same velocity $\lambda$.
Note that $\lambda < c_X$ for all $X$.  For the example we consider, in the 
event of massless
particles among the decay products, they would minimize final state energy by 
carrying exactly zero momenta \cite{CG2}.  Since the $\{c_X\}$ are the maximum
attainable velocies for the decay products, that implies that $\lambda < 
\textrm{min}\{c_X\}$.  

It is this last point we focus on: $\textrm{min}\{c_X\}$ in our 
case corresponds to $\textrm{max}\{m_X\}$.  In the ultrarelativistic limit, 
$E_A \approx p_A c_A$ and so on, and the particle with the lowest speed of 
light will carry most of the momentum (in the minimum energy case).  Further,
in this threshold case, $\lambda$ is also the velocity of the incident 
particle \cite{CG2}.  Since $\lambda \approx c_A$ in the regime we are 
considering, that means that the decay can only occur if $c_A > \textrm{min}\{c_X\}$.  
From Eq.\ \ref{delta_ab}, that implies that $m_A < \textrm{max}\{m_X\}$.

In other words, above a certain threshold, a decay may only be allowed if
at least one decay product is more massive than the decaying particle.  This 
can have serious consequences for the particle content of UHECRs, as we will
now address.

\section{The inverse cascade}

In this model, for each individual reaction there is a threshold energy $E_T$ 
above which particles can only decay if at least one of the decay products 
is more massive.  For the decay $A \rightarrow B + ...$, where $B$ 
is the most massive decay product, 
$m_B > m_A$, and $m_{...} \ll m_B$ the threshold energy goes like
\begin{equation}
E_T \sim c^3 \sqrt{\frac{m_B^2 - m_A^2}{c_A^2 - c_B^2}}.
\end{equation}

Using equation \ref{delta_ab}, this reduces to
\begin{equation}
E_T \sim c^2 \sqrt{M(m_A + m_B)} \left(\frac{h'(0)}{1}\right)^{-1/2}.
\end{equation}

Due to the large value of the Planck mass, this threshold is around the same
order of magnitude for a wide range of reactions involving particles of mass
much less than $M$.

Consider an UHECR of mass $m_0$ produced with energy $E_0 \gg E_T$. 
If normally (at energies $E \ll E_T$), particle {\bf 0} can be a decay product of 
a more massive particle {\bf 1}, then here ($E \gg E_T$), {\bf 0} can decay 
into {\bf 1}.  We ignore the lighter products that will result as well; the
substantial fraction of the incident energy will be carried by the most massive
decay product. 

Now, we have a particle of energy $E_1 < E_0$ and mass $m_1 > m_0$.  If it
can, it will also decay into a particle {\bf 2} with $m_2 > m_1$ and energy
$E_2 < E_1$.  As the decays continue into more and more massive final states,
the final energy continues to decrease, until at some point we reach a 
particle {\bf N} with energy $E_N < E_T$ and mass $m_N > m_0$.  Since we are
now below threshold, the {\bf N} particle can decay as usual into {\bf N-1}
and so on into less massive particles.

So, to summarize, this inverse cascade decay process occurs in two stages.

1.  {\bf Decay up: }Decays into more and more massive, yet less energetic particles,
 when $E > E_T$, and 

2.  {\bf Decay down:} Regular decay schemes once $E < E_T$.

\section{Pandora's Box (Newly allowed reactions)}

We saw in Section II that at sufficiently high energies particles can only decay 
if one of the decay products is more massive.  In Section III we saw how this can
produce an ``inverse cascade'' to more and more massive particle
species.  In this section we address the specifics of how this cascade occurs, by
examining which previously forbidden decays can now occur.  All of the following are
assumed to be occuring well above threshold, and we list them in several classes.

\subsection{Tree-level QED}

Here the bare QED vertex $\gamma \rightarrow Q^+Q^-$ where $Q$ is any charged
particle is allowed as a decay.  This also includes final states with extra
photons and bound states of $Q^+Q^-$, like $\gamma \rightarrow \pi^0 \gamma$.

\subsection{Tree-level weak}

It is in the weak sector that most of the interesting decays in the inverse
cascade occur, due to the flavor-changing $W$ vertex.

In this Lorentz-violating scheme, weak decays (such as that of the neutron)
can happen in reverse, allowing proton decay via $p \rightarrow n e^+ \nu$ and
flavor changing decays such as $n \rightarrow \Lambda \pi^0$.  

Not only that, decays to weak bosons become permissible, such as $\nu_l \rightarrow l^\pm W^\mp$ and 
$l^\pm \rightarrow \nu_l W^\pm$.  Included in this are decays of leptons to
heavier leptons due to virtual $W$'s.
In the quark sector, this allows decays 
such as $p \rightarrow n W^+$ and so on.  Generally, the flavor changing weak
decay $q \rightarrow q' W$ will convert all quarks to $t$ quarks.
In the absence of a fourth 
generation, a bare $t$ quark is stable.  

Since the $Z$ is massive, decays such as $\{q,l,\nu_l\} \rightarrow Z \{q,l,\nu_l\}$ 
(``$Z$-Cerenkov'') become permissible as well.  

For the bosons; $W^+ \rightarrow t \bar b$ is allowed, and $Z$-Cerenkov is possible for 
the $W$ as well (this is a bare weak vertex, as is $W \rightarrow WZZ$ and 
$W \rightarrow WZ\gamma$).  The $Z$ is unstable to the decay $Z \rightarrow t \bar t$.

Now consider the Higgs: indeed we might be above the energy
of weak symmetry breaking.  Including a single massive Higgs $H$ would
allow; $\{f,W,Z\} \rightarrow \{f,W,Z\} H$ (``Higgs-Cerenkov''
 for fermions and weak bosons) and $\{W,Z\} \rightarrow \{W,Z\} HH$.  In the
absence of new physics beyond the Standard Model, the $H$ is stable if it is heavier
than the top quark.  If it is lighter than the top, the Higgs is unstable to $H \rightarrow
t \bar t$.

\subsection{Tree-level Strong}

Since the bare QCD vertex preserves quark flavor and charge, not much happens
in the bare quark sector, apart from the reversings of strong decays such
as $N \rightarrow \Delta \pi$ and $\pi \rightarrow \rho \pi$.  Just as for
the massless photon, free gluon decay is allowed via $g \rightarrow q \bar q$.

\subsection{One-loop processes}

One could formulate many more decays at the 1-loop level, but the rates for these 
for them are suppressed compared to the tree level decays.  
For this reason, we do not consider them 
further here, except to note two interesting decays of the photon:
 
$\gamma 
\rightarrow \nu_l \bar \nu_l$ (QED and weak process).  Note that since
neutrinos are so much lighter than their heavy partners and the quarks, the 
threshold for photon to neutrino decay could be very low depending on how light
the neutrinos are.  

$\gamma \rightarrow Z \gamma$ (weak process).  This photon $Z$-Cerenkov decay
occurs due to a $W$-loop.

\subsection{Spin up}

For bound states (and now restricting ourselves to bound states of three valence quarks)
there exist higher mass resonances that only differ in the mass and total
angular momentum (e.g., the $\Delta$ and higher resonances for the proton and
neutron).  Since the state with higher angular momentum is more energetic, it 
is more massive and thus processes like $N \rightarrow N^* \pi$, $N* \rightarrow N^{**} \pi$ and so on can act to ``spin up'' the particle to more and more
massive states.

Thus, even the all top quark baryon $ttt$ isn't completely stable, but can 
decay via spin up.

\section{Estimates of decay rates}

For the photon decay $\gamma \rightarrow e^+ e^-$, the decay rate $\Gamma$ well
above threshold was 
\begin{equation}
\Gamma_{\gamma \rightarrow ee} = \frac{1}{2} \alpha \delta_{\gamma e} E.
\end{equation}

We adapt this form of the decay rate to other processes to make estimates, and argue for it as
follows.  At ultra-relativistic energies, the only energy scale is the incident
energy $E$, so $\Gamma \propto E$.  It is proportional also to the effective
coupling constant involved in the decay, and to kinematic factors.  We 
schematically write this as
\begin{equation}
\Gamma \sim (coupling)\times(kinematics)\times E.
\end{equation}
For the photon decay, the coupling is $\alpha$ and the kinematic factor is
$\delta_{\gamma e}$.  We generalize these factors as follows.  

We (first) consider decays of the form
$A \rightarrow B + C$ where $m_A < m_B$, $m_B \gg m_C$.  At threshold and
beyond, most of the momentum of the final state is carried by the particle
with the lowest speed of light, thus highest mass (particle $B$).  The relevant
kinematical factor will then be $\delta_{AB}$.  This will continue to be true
when final states with more than 2 particles are allowed, as long as $B$ continues
to be the most massive particle.  

The coupling factor is determined by looking at the tree level Feynman diagrams and
counting vertices involved in the reaction.  

Consider then the decay of the proton; in addition to the inverse of the neutron decay
$p \rightarrow n e^+ \nu_e$ (``leptonic'' decay),
the decay $p \rightarrow n W^+$ (``bosonic'' decay) is also allowed.

For the bosonic decay; $E_T \sim \sqrt{m_W M} \sim 2.8\times10^{19}$ eV, and
as most of the energy goes into the $W$,
$\delta_{pW} \sim m_W / 2 M \sim 4\times10^{-18}$.  The vertex is a 
single weak vertex; with coupling $\alpha_W = \alpha / \sin^2 \theta_W 
\sim 0.0316$.  Thus, at $E = E_T, \Gamma \sim 4$ eV.  

For the leptonic decay; $E_T \sim \sqrt{M (2 m_p)} \sim 4.5\times10^{18}$ eV,
and as most of the energy goes into the $n$, $\delta_{pn} \sim (m_n - m_p)/ 2 M \sim 5\times10^{-23}$.  
This reaction has two weak vertices with 
coupling $\alpha_W^2 \sim 10^{-3}$.  Thus, at $E = E_T, \Gamma \sim 2.2\times
10^{-7}$ eV.  

Note that the distance travelled before decay is $\sim 1/\Gamma$, and given
$\hbar c = 6.57\times10^{-24}$ eV pc, the $1/\Gamma$ for these reactions is
incredibly small: $10^{-24}$ pc for the bosonic and $10^{-17}$ pc for the 
leptonic.  In the context of a UHECR, assuming comparable rates (within a few
orders of magnitude) to the above, this inverse cascade happens 
essentially immediately, so that by the time any decay products reach the 
Earth, they have gone through both stages of the cascade and are now 
sub-threshold particles.  

\section{Consequences for UHECR composition}

Given the rapid rate of the inverse cascade process, there can be two outcomes
for the UHECR population at earth.  

First, if both the Decay Up and Decay Down 
sequences occur, then the consituent particles could, depending on the exact
decay chains and nature of initial UHE primary, consist of protons, electrons, 
photons, and their antiparticles.  Each particle spectrum would exhibit a cutoff
at their prospective threshold energies, which are all related by the $h'(0)$ parameter.

Second, and more interesting, if the initial UHE primary is of 
high enough energy, the decay up sequence may occur and leave particles that can 
no longer decay into more massive particles, but are still above threshold for decay
into less massive particles.  These Most Massive Particles (MMPs) would then be the 
most energetic component of the UHECR population.  What exactly the MMPs are depends
on the model of particle physics used.  Of particles known to exist know, the MMPs would
consist of top quarks and excited bound states of top ($t \bar t$ mesons and $ttt$ baryons)
quarks.  If the Higgs is more massive than the top, then the MMPs would consist of
single Higgs bosons.  Including Supersymmetry and its expected spectrum of more massive
superpartners, if the decay rate is still rapid enough the MMP would be the most massive
superparticle (MMSP).

Of course, in the first outcome, our proposed solution of using LI violation to solve
the photon abundance issues in top-down models is not obviously successful.  For both
outcomes, in fact, one would have to re-examine the decay process starting with the initial
decay of the supermassive relic (which is assumed to be sub-threshold, so it decays ``normally''),
and follow the decay products to see what results.  

Simulations of MMP-induced shower development would have to be performed to see
if they are consistent with actual events.  It was shown in Ref.\ \cite{albuquerque}
that {\it stable} UHE primaries of mass greater than about 50 GeV should be 
distinguishable from proton primaries by their atmospheric shower profiles.  As the
lightest conceivable MMP (a $t \bar t$ meson) would be much heavier than this
bound, the result of \cite{albuquerque} might be used to exclude the MMP as the
primary for UHE showers.  However, a key assumption is that the primary is stable. 
For a MMP just above the threshold energy, it may take only a few collisions before
the MMP energy is lowered below threshold, where it will decay ``normally'' (e.g.,
$t \bar t \rightarrow W^+ W^-$) into a shower of particles, perhaps mimicking the 
shower of a nucleus UHECR.  A MMP with incident energy farther above threshold would
then penetrate deeper.  In any event, further investigation into MMP UHECR shower
profiles is warranted.

Also of concern is that, while energy and momentum are conserved in a certain 
preferred frame, they are necessarily not conserved in all other frames.  As such, 
due to the motion of the earth and the solar system, small departures from energy 
conservation might be observed as well in the UHECR atmospheric showers.

This work was supported in part by the Department of Energy and by NASA (NAG5-10842) 
and by NSF grant PHY 00-79251. 


\end{document}